**Abstract** Use of resonant light forces opens a unique approach to high-volume sorting of microspherical resonators with much higher uniformity of resonances compared with that using the fabrication of microresonators by lithography and etching. In this work, the spectral response of the propulsion forces exerted on polystyrene microspheres near tapered microfibers is directly observed. The measurements are based on the control of the detuning between the tunable laser and internal resonances in each sphere with accuracy higher than the width of the resonances. The measured spectral shape of the propulsion forces correlates well with the whispering gallery mode resonances in the microspheres. The existence of a stable radial trap for the microspheres propelled along the taper is demonstrated. The giant force peaks observed for 20 μm spheres are found to be in a good agreement with a model calculation demonstrating an efficient use of the light momentum for propelling the microspheres.

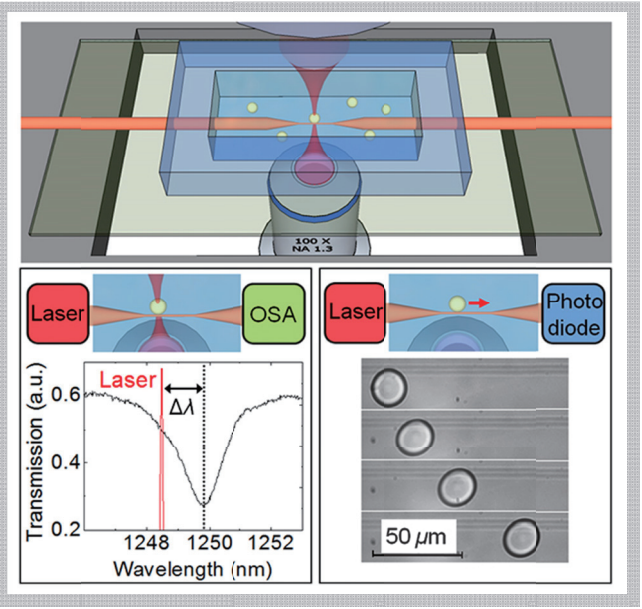

# Spectrally resolved resonant propulsion of dielectric microspheres

Yangcheng Li[1,*], Alexey V. Maslov[2], Nicholaos I. Limberopoulos[3], Augustine M. Urbas[4], and Vasily N. Astratov[1,3,*]

## 1. Introduction

In 1977 Arthur Ashkin and Joseph Dziedzic observed resonant enhancement of the optical forces exerted on microdroplets by an optical beam [1], which was arguably the first known cavity optomechanics effect. For a long time, however, the studies of this effect have been scarce because the resonant force peaks have been weak and difficult to observe [2, 3]. In recent years, the situation changed dramatically since the resonant light forces became a cornerstone of many cavity optomechanics effects including dynamic back action [4], self-induced [5, 6] and carousel [6] trapping, and resonant propulsion [7-9] of microparticles. Theoretically, stronger force peaks have been predicted for evanescent field couplers [10-15].

The knowledge of the shape of resonant features in the spectral response of optical forces is critical for understanding the mechanisms of various cavity optomechanics effects. For the structures fixed on a substrate such as coupled microring-strip waveguide systems the force spectrum displays an asymmetric Fano line shape which modifies in accordance with the phase factor [16, 17].

For free-moving particles in the vicinity of evanescent couplers, such as surface waveguides, tapered fibers, or totally internally reflecting prisms, the resonant optical forces can be used for spatially separating the particles with the desirable resonant properties [18, 19]. Recently we showed that traversing the focused laser beam by a particle in a free-fall can lead to an efficient sorting effect [19]. The principle of sorting is based on extreme sensitivity of the magnitude of the optical force to the laser wavelength detuning ($\Delta\lambda$) from the resonant wavelengths of whispering gallery modes (WGMs) in microspheres. The sorting precision is determined by an inverse of the WGM's quality factor ($1/Q$), which far exceeds the current fabrication capability of ~1% standard deviation in sphere diameters. This approach can be applied to sorting

---

[1] Department of Physics and Optical Science, Center for Optoelectronics and Optical Communications, University of North Carolina at Charlotte, Charlotte, NC 28223-0001, USA
[2] Department of Radiophysics, University of Nizhny Novgorod, Nizhny Novgorod 603950, Russia
[3] Air Force Research Laboratory, Sensor Directorate, Wright-Patterson AFB, OH 45433 USA
[4] Air Force Research Laboratory, Materials and Manufacturing Directorate, Wright Patterson AFB, OH 45433 USA
* Corresponding authors: e-mails: yli63@uncc.edu; astratov@uncc.edu; telephone: 001-704-687-8131; fax: 001-704-687-8197

microspheres made from different materials, both in liquid and in air environment [18, 19]. Microspheres having identical resonant wavelengths can be used as building blocks of coupled-cavity devices where the disorder and localization effects are strongly suppressed [20]. Examples of such devices are coupled resonator optical waveguides (CROWs) [21-23], high-order spectral filters [24], delay lines [25], sensors and microspectrometers.

From the point of view of developing such *microspherical photonics*, an interesting task is to sort high-index spheres ($n>1.9$) which can possess $Q>10^4$ in sufficiently small particles, $D\sim3$ μm, suitable for developing compact coupled-cavity devices. In particular, it can be achieved using microspheres made from chalcogenide glasses [26], barium-titanate glasses [27], lead silicate glasses [28], germanate glasses [29], telluride glasses [30], silicon [31] or other high-index materials. These high-index spheres, however, have much higher densities than water or other liquids, so that they quickly sink in a liquid that complicates their manipulation in the vicinity of evanescent couplers. For this reason, it is likely that sorting of such high-index spheres will be achieved by a modified technique [18, 19] by allowing the particle to traverse the laser beam in a vacuum or gaseous environment. Still, the mechanisms of resonant light forces are very similar in different media.

The motivation of this work is to understand the basic physics of the resonant propulsion effects, and for this reason we focus on polystyrene microspheres (with the density very similar to water) which allow developing very efficient schemes of optical manipulation of spheres in a water environment. It should be noted that the optical forces exerted on particles in a liquid are rather complicated involving: i) propelling forces along the evanescent coupler, ii) drag forces, iii) transverse forces required for stable particle separation from the surface of the coupler, iv) rotational forces.

An additional factor is represented by photophoretic forces caused by non-uniform heating of the particles. For absorbing metallic particles, the photophoretic forces can be much stronger than the optical forces [32]. For dielectric particles, however, the extremely small absorption dramatically reduces the role of photophoretic forces.

Optical propulsion of dielectric microspheres has been studied in various evanescent field couplers [33-37]. However, the resonant enhancement of the optical forces was too weak in these experiments. More recently, ultrahigh peaks of resonant forces were observed in propulsion experiments in tapered fiber couplers [7]. However, the detuning between the laser and WGMs in different spheres was varied randomly in Ref. [7] due to the inevitable 1-2% size deviations of microspheres.

In this work, we directly measured the spectral shape of the peaks of the resonant optical forces exerted on microspheres in tapered fiber couplers. It was achieved using a precise control of the detuning between the laser and WGMs in individual spheres. For small detuning ($|\Delta\lambda|<\lambda_0/Q$, where $\lambda_0$ is the average wavelength), we demonstrated the existence of a stable radial trapping of microspheres near the tapered fiber in the course of their millimeter-scale propulsion. In contrast, for larger detuning we observed the lack of radial trapping of microspheres. We demonstrate a very broad and weakly pronounced spectral peak in the force for 10 μm polystyrene spheres and strongly pronounced sharp force peak with $Q>10^3$ for 20 μm spheres. We show that the shape of the peak of the spectral response of propulsion forces replicates, with the opposite sign, the shape of the dip in fiber-transmission spectra. We observed significantly higher magnitude of the peak forces compared to previous measurements [7] and showed that this value reaches the total absorption limit ($F\sim P_0/c$) for the light momentum flux.

## 2. Two-dimensional Theoretical Model and Discussion of 3-D Case

In this Section we consider a two-dimensional (2-D) model explaining the origin of longitudinal ($F_x$) and transverse ($F_y$) forces exerted on microspheres in evanescent couplers, as schematically illustrated in the inset of Fig. 1. Our approach and results bear similarity with previous force calculations performed for different evanescent couplers [10-15]. Compared to our previous calculations performed for a wave guided by a metallic boundary [7, 8], in this work we consider a thin dielectric slab coupled to a cylindrical resonator.

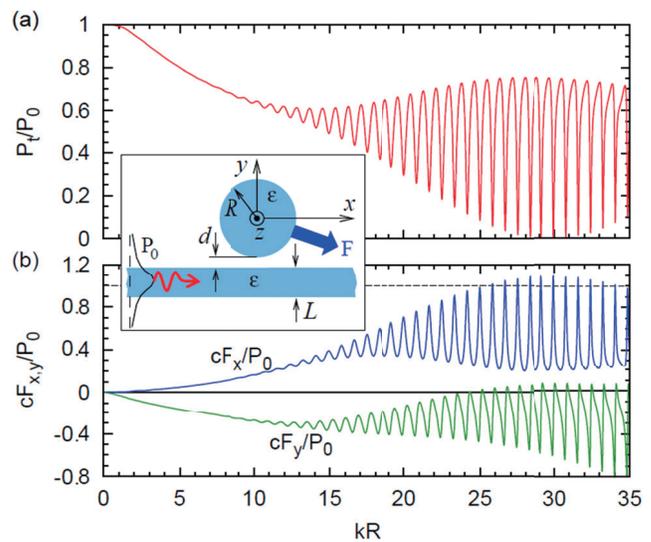

Figure 1. (a) Transmitted guided power $P_t/P_0$, (b) longitudinal, $cF_x/P_0$, and transverse, $cF_y/P_0$, forces as functions of the particle size parameter $kR$. Inset illustrate the geometry and parameters of 2-D model.

This model is a closest 2-D analog of our experiments with tapered fiber couplers. It should be noted, however, that the splitting of WGMs well-known for 3-D real physical spheres cannot be accounted for by this model. WGMs in microspheres are characterized by three modal numbers, $q$, $l$ and $m$ [20]. The radial number $q$ represents the number of the WGM intensity maxima in a radial direction. The orbital number $l$ represents the number of wavelengths along the sphere equator. The azimuthal number $m$ takes ($2l+1$) values from $-l$ to $+l$ and represents the modes effectively spreading away from the equatorial plane toward the poles of microsphere [38-42]. In a perfect sphere, the azimuthal modes are degenerate. However, in real experimental situations the degeneracy of azimuthal modes is often lifted. It can take place due to elliptical deformation of microspheres. It can also take place due to a perturbation induced by contact with the substrate or any other object in the vicinity of the sphere surface. Spectral overlap between several simultaneously excited azimuthal modes usually leads to dramatically decreased values of measured $Q$-factors of WGMs compared to single-mode calculations. This feature of 3-D geometry is outside the scope of our simplified 2-D model. The closest 3-D analogy of our 2-D calculations would be perfect spheres with degenerate azimuthal modes. For such perfect spheres the $q$ and $l$ numbers are analogous to corresponding numbers in 2-D geometry of circular resonators.

As illustrated in the inset of Fig. 1, the waveguiding slab has thickness $L$ and refractive index $n_g$. The scattering cylinder has radius $R$ and refractive index $n_s$ and is separated from the slab by distance $d$. To be specific, we choose $n_g = n_s = 1.30$, $kd = 1.5$ ($k=\omega/c$) and assume vacuum background. The Maxwell equations in 2-D can be split into two independent sets: transverse electric (TE) and transverse magnetic (TM). We consider only the TM case. The number of modes supported by the slab increases with the size parameter $kL$. For $n_g = 1.30$, the slab supports only one TM mode if $kL<3.78$. We take $kL = 2$ that gives the phase index for this mode $n_{ph} = c/v_{ph} \approx 1.074$.

The guided mode of the slab is scattered due to the presence of the resonator. The scattering results in the reduction of the transmitted power, creation of the reflected guided mode as well as of bulk radiation that escapes to the upper and lower half-spaces. The scattering is also accompanied by the creation of the electromagnetic force on the resonator. To find the scattered waves and the forces, we use the analytical theory that was previously applied to a simpler system – cylinder interacting with the surface wave of a plasma half-space [8]. The theory is based on representing the scattered fields outside the resonator using effective surface currents and the Green's function for a source located near the slab. The fields inside are expanded in terms of cylindrical functions. Matching the fields at the boundary of the cylinder gives the effective currents and the fields inside. To adopt the formulas of Ref. [8] to the present model, we replaced the corresponding Green's function with the one that describes the emission near the slab. The calculated fields can then be used to find the propelling and trapping forces using either the Lorentz formula or the Maxwell stress tensor.

The dependence of the transmitted power and forces on the parameter $kR$ is shown in Fig. 1. For small $kR$, the transmitted power decreases monotonically with $kR$ due to increased scattering of light by the cylinder. However, at $kR>15$ the transmission starts to exhibit narrow dips. The dips are due to the excitation of the first order ($q=1$) WGMs in the resonator. At $27<kR<32$, the transmission loss at the dips can reach almost 100%. This situation corresponds to a critical coupling condition when the power is coupled to WGMs in microspheres almost completely and subsequently scattered in the surrounding space. The fraction of the power reflected in the waveguide in the backward direction is too small to contribute markedly to the balance of photon fluxes. Despite the large total power of the bulk radiation, its momentum flow along the incidence direction is rather small due to its lack of directionality. Thus, on resonance with WGMs the momentum flow, $n_{ph}P_0/c$, of initial waveguide mode is transferred almost entirely to the cylinder and the normalized propelling force peaks reach the values $cF_x/P_0 \approx 1.1$ in good agreement with the results of the calculations presented in Fig. 1(b) for $27<kR<32$. The peaks positions of the propelling force, $F_x$, also correlate well with the dips in the transmission spectrum.

Unlike the propelling force that exhibits well defined resonant peaks, the transverse (trapping) force $cF_y/P_0$ has a more complicated behavior. Its spectral features are clearly resonant in nature, but highly asymmetric. A phenomenological theory shows that the force on the resonator can be decomposed into two terms: symmetric and anti-symmetric [43]. Depending on specific parameters, their combination gives an asymmetric line shape. Similar line shapes have been also reported in Refs. [11-17]. The results presented in Fig. 1(b) show that the transverse force can be attractive ($F_y<0$) and have a resonant nature. It should be noted, however, that the repulsive peaks of the optical forces have also been reported in previous studies [13]. Generally, the spectral shape, magnitude and direction of the transverse forces are sensitive functions of the structural parameters ($k$, $R$, $n_s$, $n_g$ and $d$), and require calculations in specific cases.

Most of these considerations should also be applicable in a 3-D geometry of tapered microfibers realized in the experimental Section 3 of this work. We should note, however, that in 3-D geometry of tapered fibers the non-resonant optical forces should be additionally suppressed relative to the resonant peaks. Away from the resonance, the microsphere experiences only a fraction of the total power carried by the evanescent field due to the fact that the sphere is placed from one side of the fiber. On contrary, close to critical coupling the optical power can be passed from the taper to the dielectric microsphere almost entirely. This means that in the 3-D geometry of microsphere-to-microfiber couplers we expect well pronounced negative (attractive) force peaks with weaker forces in between the peaks compared to calculations in 2-D geometry. The splitting of azimuthal WGMs discussed earlier can also lead to broadening of the peak forces in 3-D geometry due to overlap of the peak forces associated with individual azimuthal modes.

## 3. Experimental Results and Discussion

### 3.1. Experimental apparatus

The experimental setup and sequence of procedures are illustrated in Fig. 2. We selected the tapered microfiber for propulsion experiments with microspheres for several reasons. It allows fabricating extremely thin tapers (micron-scale diameter) with a large fraction of power guided as evanescent fields outside. The tapers obtained by wet etching are only few millimeters long that allow avoiding problems of their accidental breaking by fixing them in specially designed platforms [7, 27]. Finally, it allows taking advantage of fiber-integrated technology including tunable laser sources and spectrometers.

The propelling force can be determined due to the fact that the terminal velocity ($v$) of the particle is reached when the propelling force ($F_x$) is equal to the drag force, $3\pi\mu Dv$, where $\mu$ is the dynamic viscosity. Thus, the goal is to measure $v$ and calculate $F_x$. The propulsion of small ($D<10$ μm) polystyrene particles is a steady process, and the measurements of $v$ are straightforward [33-37]. For larger spheres ($15<D<25$ μm), however, the WGM resonances become more pronounced ($Q>10^3$) that results in significant increase of $v$ for some of the spheres [7]. It should be noted that the propulsion becomes less steady for larger spheres. This behavior may have different explanations including possible braking caused by touching the taper by more massive spheres or by variations in the coupling efficiency. Slight deviations of the particles from the spherical shape can play some part in this process since as the particle rotates the coupling conditions with various azimuthal modes can vary. In practice, it means that only maximal velocity of propulsion can be considered as a measure of the optical forces under conditions of unrestricted motion of spheres.

In the course of propulsion the microspheres are separated from the taper by a small liquid gap. It can be explained by a combination of long-range optical attraction (gradient force) and a short-range electrostatic repulsion caused by the same (negative) charge on the surface of the silica taper and polystyrene microparticles [6]. The size of this gap (<200 nm) was beyond the resolution of our imaging system. We were able, however, to qualitatively detect the variations of this gap in the course of propulsion, as discussed in Section 3.3.

In our previous studies, we analyzed propulsion movies frame-by-frame to find the maximal velocities of propulsion of different spheres [7]. We relied on slight variations of the diameter from sphere to sphere which translated into random variations of the detuning $\Delta\lambda$ between the laser and WGMs in the spheres, so that some of the spheres turned out to be in resonance with the laser operating at a fixed wavelength.

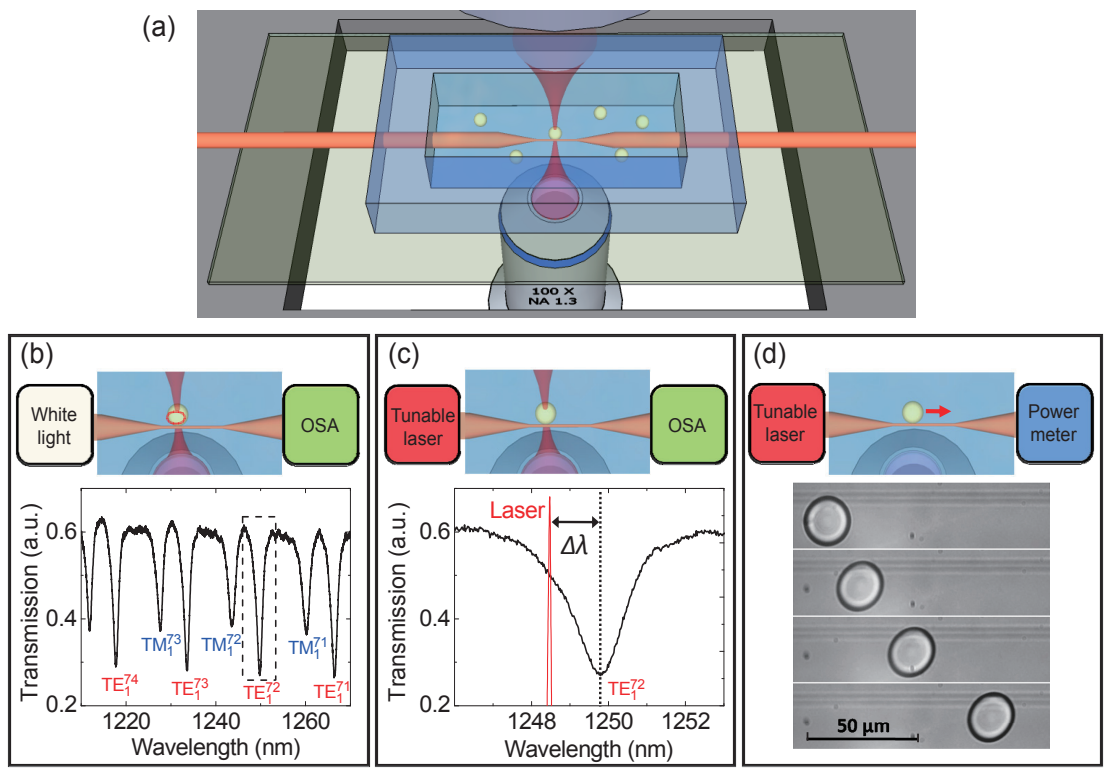

Figure 2. (a) Overview of the tapered fiber coupler illustrating the manipulation of microspheres by the optical tweezers. (b-d) The sequence of experimental procedures: (b) Characterization of WGMs in a given sphere by fiber-transmission spectroscopy, (c) Setting the detuning of the laser emission line (narrow peak), $\Delta\lambda$, from the center of the dip in transmission spectrum, (d) Propelling of the 20 μm sphere along the tapered fiber represented by the snapshots taken with 100 ms time intervals.

In this work, we developed a direct control of parameter $\Delta\lambda$ for each sphere. It required two modifications of our experimental apparatus. First was the sphere manipulation capability achieved by addition of optical tweezers to our tapered fiber platform, as shown in Fig. 2(a). Optical tweezers can precisely trap and move microspheres in an aqueous environment. In particular, an individual sphere could be trapped and brought to the vicinity of the tapered fiber. Second was the use of spectroscopic characterization of WGMs in individual spheres. For transmission measurements the fiber was connected to a broadband white light source (AQ4305; Yokogawa Corp. of America, Newnan, GA, USA) and an optical spectral analyzer (AQ6370C-10; Yokogawa Corp. of America).

The microfluidic platform sketched in Fig. 2(a) was fabricated using a plexiglas frame with a single mode fiber SMF-28 fixed in a sidewall of this frame [7, 27]. The center of the fiber was etched in a droplet of hydrofluoric acid to a waist diameter of ~1.5 μm with several millimeters in length. The frame was sealed with a 100 μm thickness glass plate at bottom and filled with distilled water. The spheres used in our experiments are polystyrene microspheres (Duke Standards 4000 Series, Thermo Fisher Scientific, Fremont, CA, USA) with refractive index of 1.57 at $1.2<\lambda<1.3$ μm and mass density of 1.05 g/cm$^3$. The choice of polystyrene microspheres is determined by their density being very close to the density of the water. This was critically important condition for observation of sustained optical propulsion along the taper. The platform was placed on an optical tweezers set-up built by an Nd:YAG laser and an oil-immersed objective of 1.3 numerical aperture (NA). White light illumination was provided from the top and imaging was realized with the same objective and a CMOS camera.

### 3.2. Experimental procedure

The sequence of our propulsion experiments is illustrated step-by-step in Figs. 2(b-d), respectively.

First, the WGMs resonances were characterized for a given sphere, as shown in Fig. 2(b). While the sphere was held by the tweezers in close vicinity to the taper, white light was coupled into the fiber and transmission spectrum was measured by the optical spectral analyzer. The spectrum in Fig. 2(b) shows two sets of first-order ($q$=1) WGMs with different polarization, $TE_q^l$ and $TM_q^l$.

Second, the detuning $\Delta\lambda$ was set, as shown in Fig. 2(c). To this end, the fiber was coupled to a single mode tunable semiconductor laser (TOPTICA Photonics AG, Grafelfing, Germany) operating in 1180-1260 nm wavelength range. The optical power in the tapered region was limited to ~18 mW. Fig. 2(c) illustrates a blue shift $\Delta\lambda$ of the laser emission line relative to the resonance dip. Since the polystyrene microspheres used in our experiments have a size polydispersity of 1-2%, the resonant wavelengths are shifted from sphere to sphere. Therefore, we selected for each sphere a well-pronounced dip for the same polarization, $TE_q^l$ in transmission spectrum and measured $\Delta\lambda$ from its center.

Third, we switched off the optical tweezers beam in attempt to release the sphere for propulsion along the taper. We found, however, that the stable propulsion was observable only for small detuning ($|\Delta\lambda|<\lambda_0/Q$, where $\lambda_0$ is the average wavelength), as illustrated in Fig. 2(d) for 20 μm sphere.

### 3.3. Radial trapping of microspheres: Existence and Stability

This Section is devoted to transverse forces which are extremely important for achieving stable propulsion along the taper. We begin with summarizing main results obtained previously in a plane-parallel geometry for spheres moving in water in the vicinity of a totally internally reflecting prism [2, 10-15]. Stable transverse trapping (perpendicular to the surface of the prism) requires a combination of long-range attraction with a short-range repulsion. The long-range attraction has been attributed to the gradient optical forces which have ~100 nm spatial extents determined by the exponential decay of the evanescent field [10-15]. In addition, gravity provided downward force bringing the spheres closer to the surface of the prism. The short-range repulsion was ascribed either to electrostatic double-layer effect with the Debye screening length of about 40 nm or to radiative pressure effects caused by the imperfection-induced scattering of light away from the dielectric interface [14].

As discussed earlier in Section 2, the transverse force calculations performed using a simplified 2-D model [10-15] are not fully applicable to 3-D microfiber geometry. Qualitatively, however, we expect a similar interplay of forces, with the peaks of attractive force on resonance with WGMs in spheres combining with short-range double-layer electrostatic repulsion to create radial trapping of microspheres in the course of their long-range propulsion [7]. This scenario, however takes place only for resonant or near-resonant conditions, $|\Delta\lambda|<\lambda_0/Q$.

The situation away from the resonance is more complicated since the transverse optical forces are weakened in the 3-D microfiber geometry case, as discussed in Section 2. For the same optical power in the taper, the depth of the radial optical trap can be reduced in non-resonant situations. It should be noted that a distinction between the resonant and non-resonant case is possible only for sufficiently large spheres with high $Q$-factors of WGMs.

Experimentally, we realized non-resonant situations using polystyrene spheres with $D$=20 μm with $Q\sim10^3$. For large detuning, $|\Delta\lambda|>\lambda_0/Q$, after releasing the spheres near the taper, we were not able to observe propulsion effects. In such cases, the spheres broke away from the taper and slowly diffuse away from it. In contrast, at the same powers limited at 18 mW under conditions of small detuning ($|\Delta\lambda|<\lambda_0/Q$) we observed stable propulsion effects.

To study the strength and stability of the radial trap in resonant cases we set the zero detuning ($\Delta\lambda=0$) and perform synchronous measurements of the propulsion velocity and the total power transmitted through the fiber in the course of the sphere propulsion. The idea of this experiment was that the transmitted power is a measure of the gap separating the sphere from the taper. In the weak-to-critical coupling regime, smaller transmitted power implies a smaller gap between the sphere and the fiber surface.

For these studies, we slightly modified the sequence of experimental procedures to synchronize the measurements of the instantaneous propulsion velocity and the optical power transmitted through the fiber. The power meter in Fig. 2 (d) was replaced by a photodiode connected to an oscilloscope. Before releasing the sphere near the taper, the tunable laser beam was blocked. As a result, after releasing the sphere it tended to slowly move away from the taper. The tunable laser beam was opened with ~0.1-0.3 s delay that translated into <0.5 μm gaps between the sphere and taper as an initial condition for the propulsion process.

Motions of spheres were recorded by the CMOS camera with the average 50 ms time intervals between the frames and exposure time per frame limited at 10 ms. The exposure time was fixed during any movie recording. By reviewing these movies frame by frame, we can calculate the instantaneous velocity of sphere moving along the taper for each frame. The main factor contributing to the experimental error was accidental variation of the interval between the frames. In this method, we are able to synchronize photocurrent and instantaneous propulsion velocity and plot them as a function of time.

The results of simultaneous measurements of the propulsion velocity and the transmitted power are presented in Fig. 3 for two representative cases illustrated in Figs. 3(a) and 3(b), respectively. They were selected from up to 50 different propulsion cases. A trigger signal threshold was set so that the oscilloscope would start recording at the same time when the laser for propulsion was turned on. An initial abrupt increase of the photocurrent is explained by switching on the tunable laser beam under conditions of weak coupling with the sphere separated by up to 0.5 μm gap. The following gradual decrease of the photocurrent on the 100-200 ms time scale is explained by reducing gap between the sphere and taper. It is seen that the propulsion velocity was gradually increasing at this stage which is explained by the increased coupling strength.

As shown in Fig. 3(a), after that a relatively stable small level of photocurrent was measured which, in combination with a constant level of high propulsion velocity, indicates that the sphere settled in a stable radial trap. It is seen that the sphere stays in a stable radial trap for ~0.3 s (in some cases up to 2-3 s) covering sub-millimeter distances (in some cases up to 1 mm).

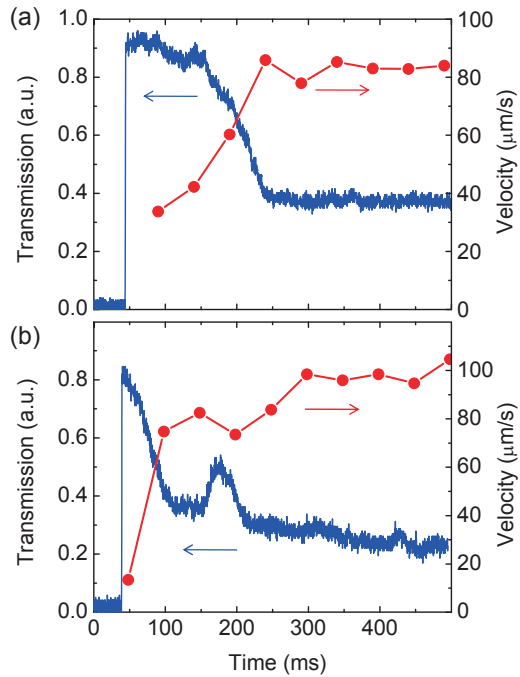

Figure 3. Fiber power transmission (photocurrent) and the propulsion velocity measured for $\Delta\lambda=0$ as a function of time for 20 μm spheres. (a) Stable radial trap achieved after 200 ms. (b) Stable radial trap is slowly approached after 500 ms.

Somewhat different scenario is illustrated in Fig. 3(b). After initial fast approaching the taper during first 100 ms, the sphere did not settle in a stable radial trap, but continued to approach the taper at a slower pace during the time interval from 100 ms to 500 ms. As expected, during this interval the propelling velocity was slowly increasing. At about 180 ms the sphere suddenly moved away from the taper as a result of fluctuation. Taper surface roughness and non-uniform charging condition in experiments may contribute to this variation. However, the depth of the transverse trap was sufficient to bring the sphere back to the taper. This case shows that only at the end of this evolution the sphere approached the stable radial trapping conditions.

### 3.4. Linearity of propulsion velocity with power

Propulsion velocity depends on many parameters of the system such as the sphere size, index and associated $Q$-factors of WGMs, the laser detuning from the WGM peaks, and the power of the beam in the tapered region of the fiber. It can be assumed that in the limit of small powers the latter dependence should be linear (for the same detuning) since the terminal velocity ($v$) is proportional to the total momentum flux. In this case, the power dependence can be excluded from the analysis by normalizing the velocity by the laser power.

We checked the power linearity of the propulsion effects for various conditions. An example of such study for polystyrene spheres with 10 μm mean diameter and zero detuning is presented in Fig. 4.

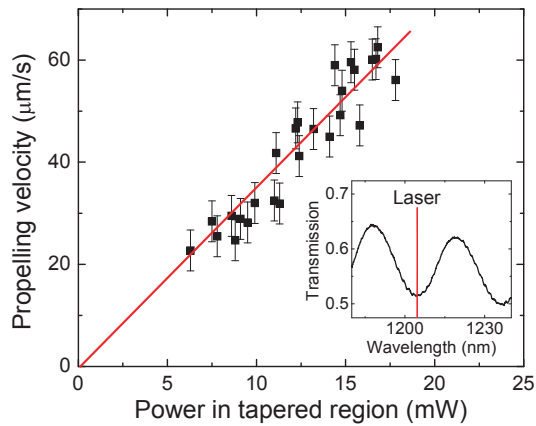

Figure 4. Optical power dependence of propelling velocity for 10 μm spheres with a linear fit indicated by a line. Insert: typical transmission spectrum for 10 μm sphere in water and laser emission line tuned at resonance dip (zero wavelength detuning).

The WGM-defined dips in the fiber-transmission spectra are very broad and shallow, as shown in the inset of Fig. 4. For each sphere the laser emission peak was tuned to the middle of one of the spectral dips. The propulsion velocity was found to be linearly dependent on the laser powers within 6-18 mW range. For smaller powers the sphere could not be retained near the taper due to insufficient depth of the radial trap.

## 3.5. Dependence of the propulsion velocity on detuning

In this Section, we present the results of measurements of propelling velocity as a function of the laser wavelength detuning ($\Delta\lambda$) relative to a center of given resonance dip in fiber-transmission spectra. The velocity measurements were performed after locking the sphere in a stable radial trap, as illustrated in Fig. 3. The measured velocities (μm·s$^{-1}$·mW$^{-1}$) were normalized by the total optical power guided in the tapered region.

The propulsion velocity measurements for polystyrene spheres of 10 μm and 20 μm mean diameters are summarized in Figs. 5(a) and 5(b) respectively. Multiple points for the same detuning represent different spheres. The coupling between a tapered fiber and a 10 μm sphere in water is weak, indicated by the broad resonance with a shallow depth seen in Fig. 5(a). For the entire range of detuning from -15 nm to +15 nm, the radial trapping followed by the optical propulsion could be observed. The measured velocities were found to be within 2-5 μm·s$^{-1}$·mW$^{-1}$ range. The spectral shape of the propulsion peak replicates the shape of the dip in the fiber-transmission spectrum with the opposite sign, as illustrated in Fig. 5(a).

The case of water-immersed 10 μm polystyrene microspheres represents effectively a transition from non-resonant optical propulsion studied previously for smaller spheres [33-37] to resonant optical propulsion observed for larger spheres with 15-20 μm diameters [7]. This can be understood by relating our experimental observations to the modeling results presented in Fig. 1, if we take into account smaller index contrast for water-immersed spheres (1.57/1.33=1.18) compared to that (1.30) used in modeling.

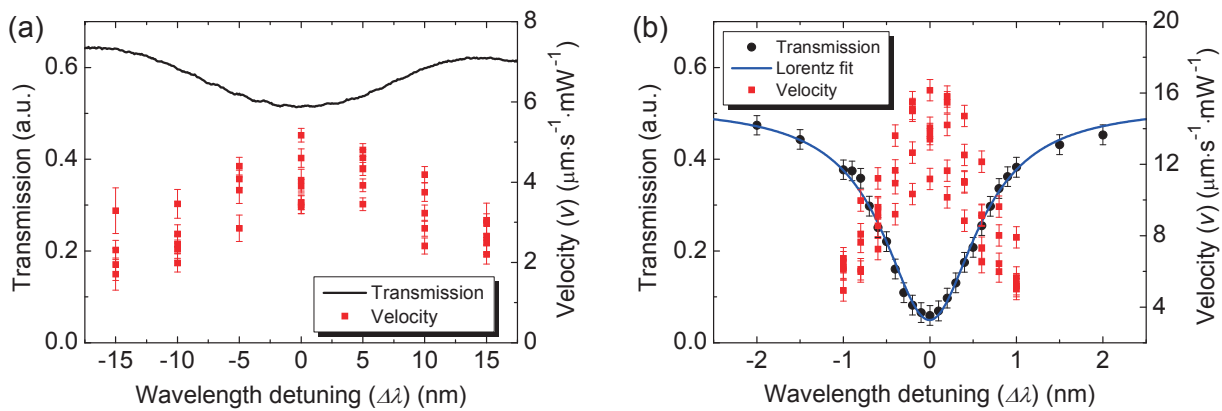

Figure 5. Transmission of the fiber with the water-immersed polystyrene microsphere positioned at the waist of the taper (solid lines) and maximal propelling velocities ($v$) normalized by the power in the tapered region (squares) measured as a function of the laser detuning ($\Delta\lambda$) from the center of the WGM-defined resonance for spheres with different mean diameters: (a) 10 μm and (b) 20 μm. For each detuning the velocity measurements were repeated on different spheres leading to multiple data points represented by squares. The curve in (b) represents a Lorentz fit to the experimental transmission measurements.

For $\lambda_0$=1.25 μm, one can estimate $kR$~25 in our experimental situation. Additional calculations for the 1.18 index contrast (not presented in Fig. 1) showed extremely weak oscillations of both forces, $F_x$ and $F_y$, for $kR \leq 25$ in agreement with the previous experiments [33-37] performed on spheres with $D$<10 μm. The force oscillations tend to increase with $kR$, however they remain weak at $kR$=25 in qualitative agreement with the results for $D$=10 μm spheres in Fig. 5(a).

In contrast, the case of water-immersed 20 μm polystyrene microspheres represents the value of $kR$~50 where the dips in fiber transmission are extremely well pronounced demonstrating $Q$~$10^3$. Calculations for the 1.18 index contrast predict strong peak forces for $kR$~50. The magnitude of the force peaks in this case is comparable to that calculated for the 1.30 index contrast in the $27 \leq kR \leq 32$ range presented in Fig. 1(b).

Experimentally, the coupling to 20 μm spheres was found to be rather close to a critical coupling regime, as illustrated in Fig. 5(b). We utilized the same tunable laser for the spectral transmission and propulsion experiments. The transmission of the fiber coupled to the microsphere was normalized by the fiber transmission without the microsphere, as illustrated by circles in Fig. 5(b). The Lorentz fit to the transmission dip with 90% (10 dB) depth and the 1.1 nm width is also illustrated in Fig. 5(b).

The propulsion velocity measurements performed on multiple spheres as a function of detuning are presented in Fig. 5(b). An extraordinary high velocity of 16 μm·s$^{-1}$·mW$^{-1}$ was observed when the laser was tuned exactly in the center of the dip in the transmission spectrum. This peak value exceeds previous measurements of non-resonant propulsion velocities in various couplers [33-37] by more than an order of magnitude. Although the resonant enhancement of the propulsion velocities was observed in our previous studies, the peak velocity value in Fig. 5(b) exceeds our previous measurements for 20 μm spheres [7] by the factor of 1.6. It can be explained by the fact that in present work we were able to set the zero detuning precisely for each sphere, whereas in our previous studies we relied on random detuning between the fixed laser wavelength and WGMs in different spheres. The distribution of the velocity over the limited detuning range, $|\Delta\lambda|<\lambda_0/Q$, was found to be inversely proportional to the shape of the dip in the transmission spectrum. As stated previously, the propulsion did not occur for spheres detuned by more than 1 nm from the center of WGM resonance because of the loss of the radial trapping for a given power.

A simple way of estimating propelling forces ($F_x$) is based on assumption that the propelling force is balanced by the drag force, $3\pi\mu Dv$. It is convenient to plot the propelling force in units of $P_0/c$ which is equivalent to the momentum flux of the incident radiation in vacuum. The momentum flux for a guided mode is larger than $P_0/c$ by an additional factor which is equal to the value of the phase index of the mode $n_{ph} = c/v_{ph}$, where $v_{ph}$ is the phase velocity of the mode. However, the convenience of not including the phase index into the normalization constant when plotting the experimental data comes from the Abraham-Minkowski controversy in which the index of the medium appears either in the numerator or denominator of the photon momentum definition [44, 45]. For the modes guided by the tapered fiber, the phase index lies between the refractive index of the taper material (silica glass, $n$=1.45) and that of the surrounding material (water, $n$=1.33). In the limit of a very thin taper, the phase index should be slightly larger than that of water and, therefore, the value $cF_x/P_0$=1 should correspond to $1/1.33 \approx 75\%$ of the conversion efficiency of the incident photon flux to the force. The resultant propelling forces (in units of $P_0/c$) are presented in Fig. 6 for 10 and 20 μm spheres.

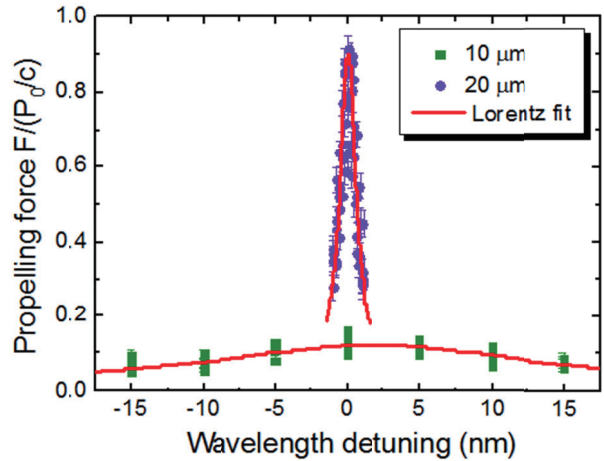

Figure 6. Comparison of optical propelling force for 10 μm and 20 μm spheres as a function of the laser wavelength detuning from the WGM resonance.

For 10 μm spheres the force peak is broad and weakly pronounced with the magnitude limited at $cF_x/P_0$~0.15. On the other hand, the peak force values for 20 μm spheres correspond to $cF_x/P_0$~0.95. As was discussed in Section 2, this is a result of very efficient coupling with WGMs in microspheres. It should be noted that observation of such extremely high efficiencies provides strong support for our model of the observed effects.

## 4. Conclusions

This work provides a direct experimental proof of existence of ultrahigh resonant propulsion forces in microspherical photonics. In contrast to our previous work [7], we took full control of propulsion experiments. It was achieved by manipulation of spheres using optical tweezers and by setting the amount of detuning between the laser and WGMs for each sphere individually. As a result, we measured the shape of the spectral response of the propulsion forces and established that it correlates precisely with the coupling properties of the microspheres. We also proved that the amplitude of the peak forces is in agreement with the model calculations.

The sorting technology developed in this work allows selecting building blocks of various photonic structures with extraordinary uniform resonant properties. In some sense, classical "photonic atoms" can behave in these applications similar to indistinguishable quantum mechanical atoms. The photonic dispersions can be engineered in such structures based on tight-binding approximation leading to applications in coupled resonator optical waveguides [21-23], high-order spectral filters [24], delay lines [25], laser-resonator arrays, spectrometers and sensors.

From the point of view of fundamental studies, particularly interesting structures are represented by large-scale 2-D and 3-D arrays of spheres where WGMs are coupled on a massive scale. The optical transport through such coupled-cavity networks is very complicated and poorly studied process which involves multiple paths for photons leading to interference, localization [46] and percolation [47] of light. The optical gain in such structures can be realized by doping the spheres with dye molecules or active ions. The combination of gain, scattering loss and optical nonlinearity in such coupled-cavity networks can result in unidirectional propagation properties [48].

By using tunable laser source it should also be possible to select spheres resonating at different wavelengths. They can be assembled in structures where both position of spheres and their individual resonant properties are controlled according to a certain design. This opens a way to explore quantum-optics analogies in photonics [49] and to create novel structures such as parity-time synthetic lattices [50], coupled resonator ladders [51] or waveguides based on an effective gauge field for photons [52].

It is likely that the applications of such structures and devices will be developed for high-index ($n>1.9$) microspheres which can possess $Q\sim10^4$ in sufficiently small particles $D\sim3$ μm suitable for developing compact coupled-cavity devices. Although the specific details of the sorting setup (sorting in vacuum, air or liquid) can vary depending on the properties of microspheres [18, 19], the present work shows the basic principles of resonant light forces useful for developing such applications.

**Acknowledgements.** The authors gratefully acknowledge support from U.S. Army Research Office through Dr. J. T. Prater under Contract No. W911NF-09-1-0450 and DURIP W911NF-11-1-0406 and W911NF-12-1-0538. Also, this work was sponsored by the Air Force Research Laboratory (AFRL/RXD) through the contract with UES, Inc.